\documentclass[aps,prl,twocolumn,floatfix,superscriptaddress,longbibliography]{revtex4-1}

\usepackage{bm}
\usepackage[pdftex]{graphicx}
\usepackage{dcolumn}
\usepackage{amsmath}
\usepackage{amssymb}

\begin{document}

\title{ On temperature-dependent anisotropies  of\\ upper critical field and London penetration depth }

\author{V. G. Kogan}
\email{kogan@ameslab.gov}
\affiliation{The Ames Laboratory and Department of Physics and Astronomy
   Iowa State University, Ames, IA 50011}

\author{R. Prozorov}
\email{prozorov@ameslab.gov}
\affiliation{The Ames Laboratory and Department of Physics and Astronomy
   Iowa State University, Ames, IA 50011}

 \author{A. E. Koshelev}
\email{koshelev@anl.gov}
\affiliation{Materials Science Division, Argonne National Laboratory,  9700 South Cass Ave., Lemont, Illinois, 60439}

  \begin{abstract}
 We show on a few examples of one-band materials with spheroidal Fermi surfaces and anisotropic order parameters  that anisotropies $\gamma_H$ of the upper critical field and $\gamma_\lambda$ of the London penetration depth depend on temperature, the feature commonly attributed to multi-band superconductors. The parameters $\gamma_H$ and $\gamma_\lambda$ may have opposite temperature dependencies or may change in the same direction  depending on Fermi surface shape  and on character of the gap nodes.
 For two-band systems, the behavior of anisotropies is affected by the ratios of bands densities of states, Fermi velocities, anisotropies, and order parameters. We investigate in detail the conditions determining the directions of temperature dependences of the two anisotropy factors.
  \end{abstract}

 \date{3 July 2019 }
 %

\maketitle

 \section{Introduction}

In many  superconductors   anisotropy parameters of the upper critical field, $\gamma_H= H_{c2,ab }/H_{c2,c}$, and  of the London penetration depth, $\gamma_\lambda= \lambda_c/\lambda_{ab }$, are not equal. Moreover, they may have different temperature dependencies. In conventional one-band   s-wave materials, these parameters for a long time were considered the same and $T$ independent. MgB$_2$ is a good example of the current situation:  $\gamma_H(T)$   decreases on warming, whereas $\gamma_\lambda$ increases  \cite{Budko,AngstPRL02,LyardPRB02,RydhPRB04,Carrington}.
Up to date, the $T$ dependence of the anisotropy parameters   is considered by many   as caused by a multi-band character of materials in question with the common reference to   MgB$_2$.
In this work we develop an approximate method to evaluate   $\gamma(T)$ that can be applied with minor modifications to various situations of different order parameter symmetries and Fermi surfaces, two bands included.
In particular, we show that even in the one-band case $ \gamma$'s and their $T$ dependencies may differ if the Fermi surface is not a sphere or the order parameter $\Delta$ is not pure s-wave.
Our conclusions challenge  the common belief that temperature dependence of $\gamma $ is always related to  multi-band topology of Fermi surfaces.

We focus   on the clean limit for two major reasons. Commonly after discovery of a new superconductor, an effort is made to obtain as clean single crystals as possible since  those provide a better chance to study the underlying physics. A  proper description of scattering in multi-band case would have lead  to a multitude of scattering parameters which cannot be easily controlled or separately measured. Besides, in general, the scattering suppresses the anisotropy of $H_{c2}$, the central quantity of interest in this work.

To begin, it is worth recalling that the problem of the second-order phase transition at the upper critical field $H_{c2}(T)$ has little  in common with the problem of a weak field penetration into superconductors, thus {\it a priori} one should not expect $\gamma_H=  \gamma_\lambda $. Still, at the critical temperature $T_c$ the Ginzburg-Landau (GL) theory   requires $\gamma_H(T_c)=  \gamma_\lambda (T_c)$ because both are described in terms of the same  mass tensor.

At $H_{c2}(T)$,  the order parameter $\Delta(\bm r, T,\bm k_F)=\Psi({\bf r},T)\,\Omega
({\bf k}_F)\to 0$, and  Eilenberger equations for weak coupling superconductors can be linearized. The factor  $ \Omega({\bf k}_F)$  here describes the $\bm k_F$ dependence of $\Delta$ and is normalized so that the average over the full Fermi surface $\langle\Omega^2\rangle=1$.

At the second-order phase transition at $H_{c2}$,
the basic self-consistency equation of the theory can be written as  \cite{ROPP}:
\begin{equation}
   \Psi \ln \frac{T}{T_c}= \int_0^{\infty}d\rho\, \ln\tanh\frac{\pi T\rho}{\hbar}\,
 \left \langle\Omega^2 {\bm v}{\bm \Pi}\,e^{-\rho {\bm v}\cdot {\bm
\Pi}}\Psi\right\rangle
\,.  \label{selfcons3}
\end{equation}
Here, ${\bm v}$ is the Fermi velocity, ${\bm  \Pi}\!=\!\nabla +2\pi i{\bm
A}/\phi_0$, ${\bm  A}$ is the vector potential, and $\phi_0$ is the flux quantum.
In comparison to the traditional form of this equation involving sums over the Matsubara frequencies, this form contains only integrals relatively easy to deal with.

Typically, the temperature dependences of the anisotropy factors are monotonic. Therefore, to determine the direction of these dependences, it is sufficient to evaluate anisotropy values at the transition temperature and at $T\!=\!0$.

\subsection{$\bm{\gamma_H }$ and $\bm{\gamma_\lambda }$ at $\bm{T_c}$}

In this domain, the gradients $\Pi\sim \xi^{-1}\to 0$ ($\xi$ is the order of magnitude of the coherence length), and one can keep
in the expansion of  $\exp(-\rho{\bm v}  {\bm
\Pi})$ in the integrand (\ref{selfcons3}) only the linear term to obtain:
\begin{equation}
  -\Psi \delta t=  \frac{7\zeta (3)\hbar^2}{16\pi^2T_c^2}\,
   \langle\Omega^2 ({\bm v}\cdot {\bm
\Pi})^2\Psi\rangle \,,\quad \delta t=1-T/T_c. \label{GL'}
\end{equation}
This is, in fact, the anisotropic version of linearized  GL equation
$- \xi^2_{ik}  \Pi_i\Pi_k \Psi =\Psi$
with \cite{Gorkov}
\begin{equation}
  \xi^2_{ik}  = \frac{7\zeta (3)\hbar^2}{16\pi^2T_c^2 \delta t}\,
   \langle\Omega^2   v_iv_k \rangle \,.     \label{xi}
\end{equation}
The anisotropy parameter for uniaxial materials then readily follows:
 \begin{equation}
\gamma_H ^2(T_c) =\left(\frac{H_{c2,a}}{H_{c2,c}}\right)^2=
\frac{\xi_{aa}^2}{\xi_{cc}^2}=
\frac{\langle \Omega^2 v_a^2\rangle }{\langle \Omega^2 v_c^2\rangle}\,.
\label{anis_clean}
\end{equation}
As mentioned above,  the anisotropy of the London penetration depth $\gamma_\lambda $ at $T_c$ is the same as that of $H_{c2}$:
\begin{equation}
\gamma_\lambda^2(T_c)= \frac{\lambda^2_{c }}{\lambda^2_{ab}}= \gamma_H ^2(T_c) =
\frac{\langle \Omega^2 v_a^2\rangle }{\langle \Omega^2 v_c^2\rangle}\,.
\label{anis_lamTc}
\end{equation}
This can be proven also directly using a general expression for the tensor $\lambda^{-2}_{ik}(T)$, see e.g. \cite{gam-lam0}.  Since $\gamma_\lambda(T_c)$ always coincides with $\gamma_H (T_c)$, in further considerations we drop the subscript, i.e.,  $\gamma(T_c)=\gamma_\lambda(T_c)=\gamma_H (T_c)$.

 In the following we adopt Fermi surface as a spheroid with the symmetry axis $z$.  The Fermi surface average of a function $A( \theta,\phi)$   is evaluated in the Appendix A:
\begin{eqnarray}
\langle A( \theta,\phi) \rangle
&=&\frac{\sqrt{\epsilon}}{4\pi }
\int  \frac{A( \theta,\phi)}{ \Gamma(\theta,\epsilon)^{3/2}  }\sin\theta d\theta\,d\varphi\,,\qquad \nonumber\\
\Gamma&=&\sin^2\theta +\epsilon\cos^2\theta\,.
\label{average}
\end{eqnarray}
Hear  $\epsilon$ is the squared ratio of spheroid axes, $\theta$ and $\varphi$ are the polar and azimuthal angles, respectively.

\subsection{ $\bm{\gamma_H }$ and $\bm{\gamma_\lambda }$   at $\bm{T=0}$}

The low-temperature orbital $H_{c2}$ determined by Eq.\,(\ref{selfcons3}) can be approximately evaluated using variational approach with the anisotropic lowest Landau level wave function as a trial, see Appendix \ref{App-var}. This gives the low-temperature anisotropy factor
\begin{equation}
\gamma_{H}(0)\approx \gamma_{o}\exp\left[-\left\langle \Omega^{2}\ln\left(\frac{v_{x}^{2}+\gamma_{o}^{2}v_{z}^{2}}{v_{x}^{2}+v_{y}^{2}}\right)\right\rangle \right],\label{eq:AnisVar}
\end{equation}
where the optimal variational parameter $\gamma_{o}$ has to be evaluated from equation
\begin{equation}
\left\langle \Omega^{2}\frac{v_{x}^{2}-\gamma_{o}^{2}v_{z}^{2}}{v_{x}^{2}+\gamma_{o}^{2}v_{z}^{2}}\right\rangle =0.\label{eq:OptAnis}
\end{equation}

According to Ref.~\cite{gam-lam0}, the anisotropy of the penetration depth at $T=0$ in the clean case is given by
 \begin{equation}
\gamma_\lambda^2(0)=
\frac{\langle  v_x^2\rangle }{\langle  v_z^2\rangle}\,.
\label{anis_lam0}
\end{equation}
Neither gap nor its anisotropy enter this result, i.e. $\gamma_\lambda (0)$ in the clean case depends only on the shape of the Fermi surface. The physical reason for this is the Galilean invariance of
the superflow in the absence of scattering: looking at the superflow in the  frame attached to a moving element, one sees  all charged particles taking part in the supercurrent independently of their energy spectrum, so that the penetration depth $\lambda(0)$ depends only on the total carriers density.

\section{Single-band anisotropies}

 \subsection{$\bm {\Omega=1}$, isotropic s-wave.}

 We start with the simplest case of a constant gap on a Fermi spheroid. In this case  $\gamma_\lambda^2=\left\langle v_x^2     \right\rangle/\left\langle v_z^2     \right\rangle$  both  at  $T=0$ and   $T_c$. With the help of Appendix A, we find
 \begin{eqnarray}
  \left\langle v_x^2     \right\rangle =v_{ab}^2\frac{\sqrt{\epsilon}}{4\pi}\int\frac{\sin^3\theta\cos^2\varphi}{\Gamma^{5/2}}d\theta\,d\varphi= \frac{v_{ab}^2}{3}\,,\nonumber\\
  \left\langle v_z^2 \right\rangle =v_{ab}^2\frac{ \epsilon^{5/2}}{4\pi}\int\frac{\sin\theta\cos^2\theta}{\Gamma^{5/2}}d\theta\,d\varphi= \frac{v_{ab}^2\epsilon}{3}\,.
  \label{<vx2>}
\end{eqnarray}
Thus,   $\gamma_\lambda=1/\sqrt{\epsilon}$ and it is $T$ independent.
In particular, this means that $\gamma_H(T_c)=1/\sqrt{\epsilon}$ as well. In fact, this is just GL results: at $T_c$,  $\gamma_\lambda=\gamma_H= \sqrt{m_c/m_{ab}}$.

To find $\gamma_H(0)$ we note that the trial lowest Landau wave function with $\gamma_o=1/\sqrt{\epsilon}$ is actually exact solution of Eq.\ \eqref{selfcons3} for the in-plane field orientation. In this case Eq.~\eqref{eq:AnisVar} in spherical coordinates $\theta,\varphi$ takes the form:
\begin{eqnarray}
\gamma_H(0)&=& \frac{1}{\sqrt{\epsilon}} \exp\left[ - \frac{\sqrt{\epsilon}}{4\pi}
\int    \frac{\sin\theta d\theta d\varphi }{\Gamma(\theta)^{3/2}}   \ln \left(\cos^2\varphi +\epsilon\cot^2\theta  \right) \right]\nonumber\\
  \label{<gH00>}
\end{eqnarray}
One can show that the double integral here vanishes meaning that $ \gamma_H(0)=1/\sqrt{\epsilon}$.
Thus, for isotropic s-wave, temperature independent anisotropies are $\gamma_\lambda=\gamma_H=1/\sqrt{\epsilon}$.

\subsection{ $\bm{\Omega= \Omega_0\cos 2\varphi}$, d-wave on Fermi spheroid}

If the spheroid symmetry axis coincides with $z$ of the d$_{x^2-y^2}$  order parameter,
the standard normalization gives  $\Omega_0^2=2$.
Since zero-$T$ anisotropy of $\lambda$ is independent of the order parameter symmetry, we have  $\gamma_\lambda(0) =1/\sqrt{\epsilon}$, the same   as for s-wave. This is the case of all examples considered in this section.
At $T_c$  we have:
 \begin{eqnarray}
 u_x^2= \left\langle \Omega^2v_x^2   \right\rangle =  \frac{\Omega_0^2v_{ab}^2}{6}\,,\quad
  u_z^2= \left\langle \Omega^2v_z^2  \right\rangle =  \frac{\Omega_0^2v_{ab}^2}{6}\epsilon\,\qquad
  \label{<ux2>}
\end{eqnarray}
and, according to Eq.\,(\ref{anis_lamTc}), $\gamma_\lambda(T_c) =1/\sqrt{\epsilon}$. Hence, $\lambda$-anisotropies for s- and d-wave symmetries are the same, that is somewhat surprising.

\begin{figure}[tb]
\begin{center}
 \includegraphics[width=8cm]{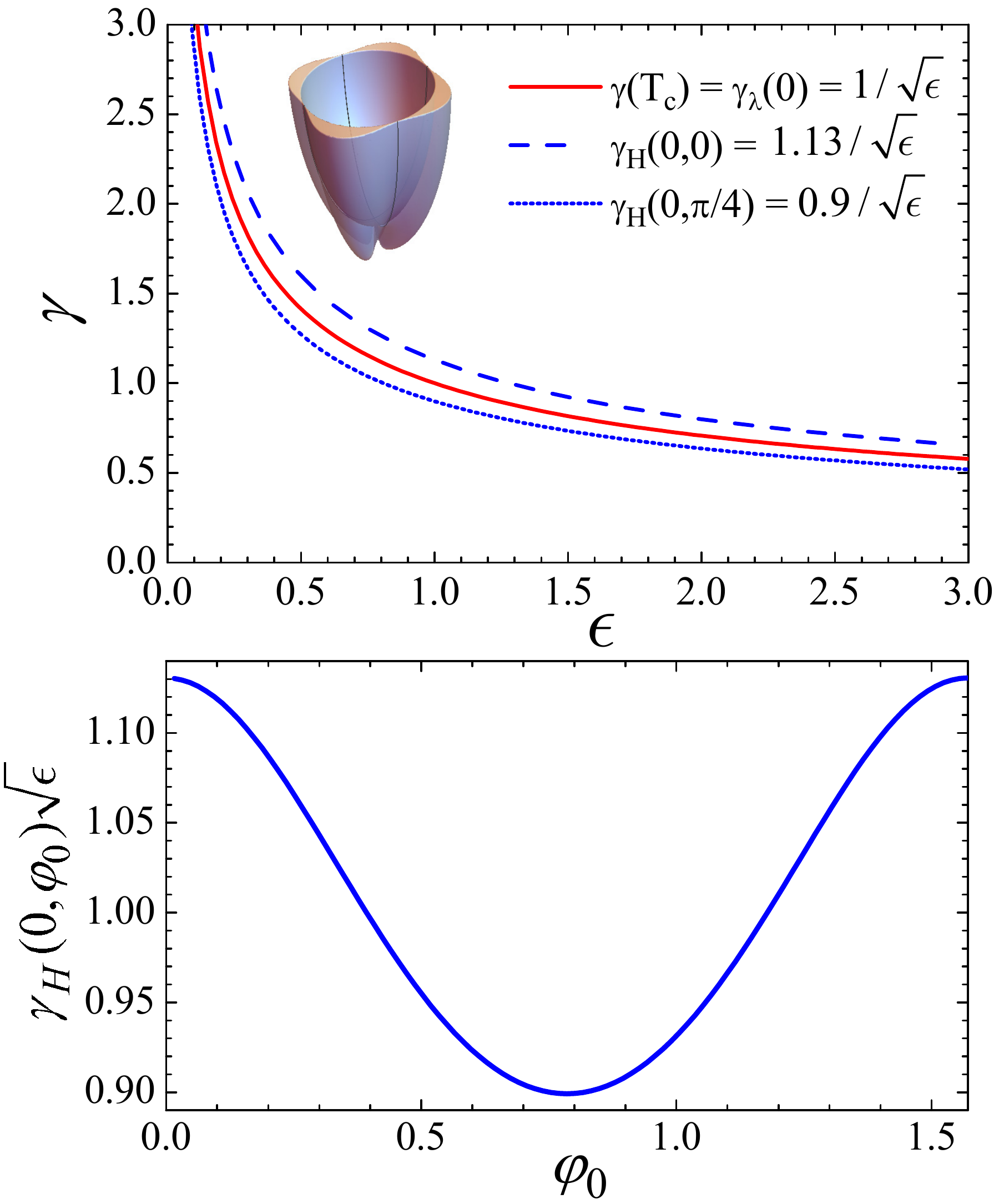}
 \caption{(Color online) \emph{The upper panel:}   d-wave anisotropies   $\gamma$ vs. ellipticity $\epsilon $.  The red curve is  $\gamma(T_c)=\gamma_\lambda(0)=1/\sqrt{\epsilon}$. The blue curves are
 $\gamma_H(0,\phi_0)$ for the in-plane $ H_{c2,y}$ parallel to the direction of the order-parameter maximum ($\phi_0=0$) shown as dashed line, and parallel to the nodal direction  ($\phi_0=\pi/4$) shown as dotted line. The inset illustrates
 the behavior of $\Omega^2$ at the Fermi spheroid.
 \emph{The lower panel} shows the angular dependence $\gamma_H(0,\varphi_0)\sqrt{\epsilon}$.
}
\label{f1new}
\end{center}
\end{figure}

The calculation of the low-temperature $H_{c2}$ anisotropy from Eqs.\ \eqref{eq:AnisVar} and \eqref{eq:OptAnis} is more involved.
In addition, with decreasing temperature the in-plane upper critical field acquires dependence on the azimuth angle $\phi_0$ between the field and direction of the maximal order parameter. Correspondingly, the $H_{c2}$ anisotropy factor also has such dependence. We will keep the $y$ axis along the direction of magnetic field, meaning that the weight function  $\Omega(\varphi)$ in the angular averaging has to be modified as $\Omega(\varphi)=-\sqrt{2}\cos[2(\varphi-\varphi_0)]$.
The angular averages in  Eqs.\ \eqref{eq:AnisVar} and \eqref{eq:OptAnis} can be done analytically.
The results, however, are somewhat cumbersome and presented in the Appendix \ref{App-dwave}.
The computed dependences of $\gamma_H(0,\varphi_0)$ on the ellipticity $\epsilon$ are shown
in the upper panel of Fig.\,\ref{f1new}. The blue dashed curve is  $\gamma_H(0,0)$, the dotted-blue is $\gamma_H(0,\pi/4)$, and the red solid curve is $\gamma(T_c)=1/\sqrt{\epsilon}$. Hence, $\gamma_H(0,0)$ decreases on warming by 13\%, whereas $\gamma_H(0,\pi/4)$ increases by 10\%. We also note that the $H_{c2}$ anisotropy at $\varphi_0=\pi/8$ is temperature independent.
At the lower panel we show the angular dependence of the product $\gamma_H(0,\varphi_0)\sqrt{\epsilon}$ which does not depend of $\epsilon$.  Hence, one expects the in-plane $H_{c2}$ to vary by about 25\% being rotated relative to the  $c$ crystal axis leading to the same variation of the  $\gamma_H(0,\varphi_0)$.

\subsection{ $ \Omega=\Omega_0\cos \theta $, an equatorial line node }
\begin{figure}[tb]
\begin{center}
 \includegraphics[width=8cm] {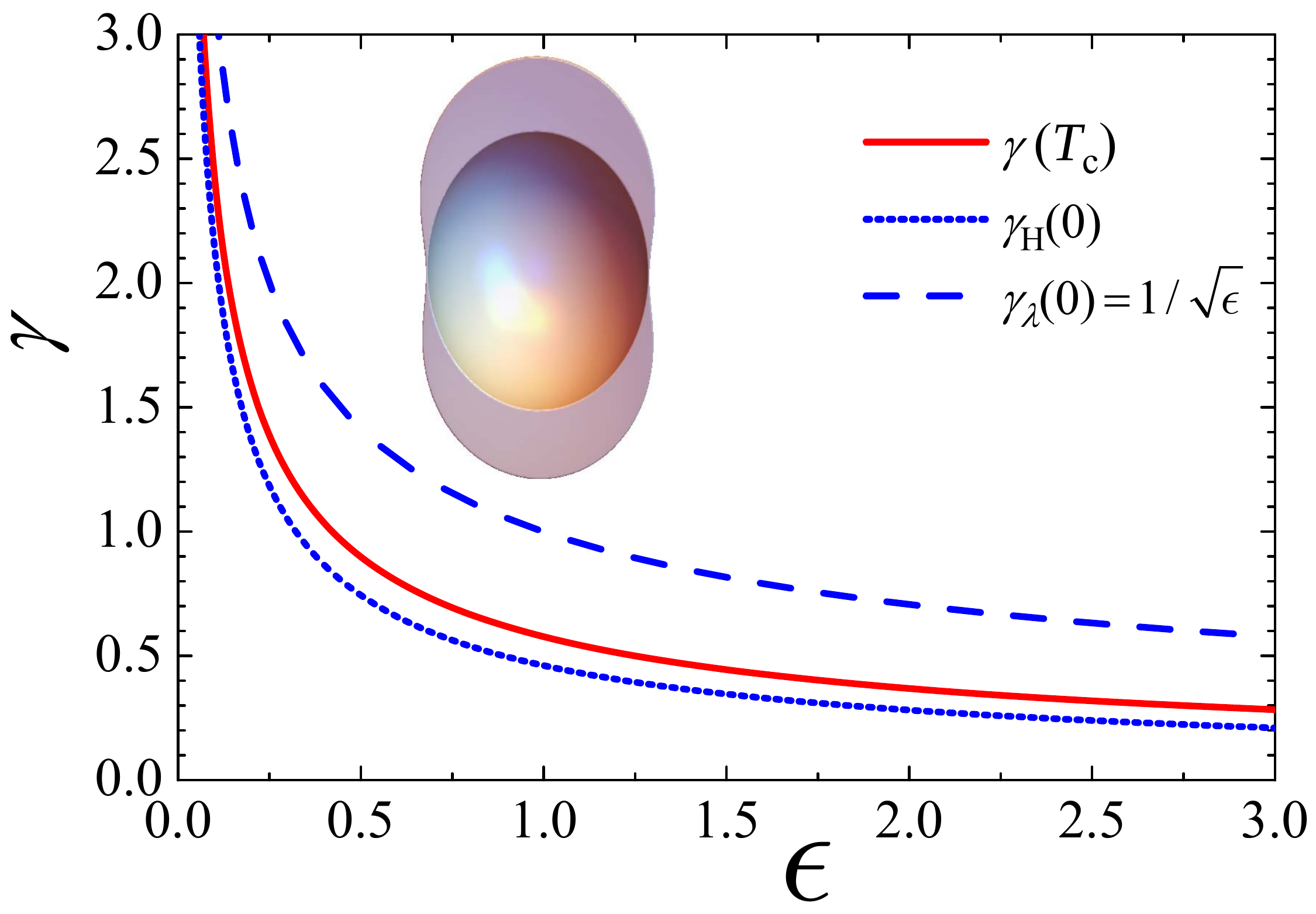}
 \caption{(Color online)    $\gamma (\epsilon)$ for the order parameter with equatorial line node.     The red curve is  $\gamma(T_c)$. The blue dashed curve is  $\gamma_\lambda(0)=1/\sqrt{\epsilon}$. The dotted-blue is $\gamma_H(0)$.
 }
\label{EqLineNode}
\end{center}
\end{figure}

The equatorial nodal line is a feature for one of realizations of the p-wave order parameter.
Again,   at $T=0$   one has  $\gamma_\lambda(0) =1/\sqrt{\epsilon}$.
The anisotropy factor at $T_c$ is
\begin{equation}
\gamma^2(T_c)=\frac{  \left\langle \Omega^2v_x^2     \right\rangle}{ \left\langle \Omega^2v_z^2     \right\rangle} =\frac{\int_0^1x^2(1-x^2)\Gamma^{-5/2}dx}{2\epsilon^2 \int_0^1x^4 \Gamma^{-5/2}dx} \,,
  \label{gam_lamTc}
\end{equation}
where $x\!=\!\cos\theta$ and $\Gamma(x)\!=\!1\!-\!(1\!-\!\epsilon)x^2$.
Integrations here can be done analytically and the result is presented in Appendix \ref{App:EqNodeLine}.
The computed anisotropy factor at $T_c$ is plotted as the solid-red curve in Fig.\,\ref{EqLineNode}.
We can see that the shape of order parameter reduces this factor in comparison with the Fermi-surface anisotropy which is represented by  $\gamma_\lambda(0)$ (blue dashed line).
As a consequence,  $\gamma_\lambda $ decreases substantially on warming.

For evaluating $\gamma_H(0)$, we need to compute averages in Eqs.~\eqref{eq:AnisVar} and \eqref{eq:OptAnis}. The normalization constant from the condition $\Omega_0^2\langle\cos^2\theta\rangle=1$ can be found as
\begin{equation}
 \Omega_{0}^{2} =\frac{(1-\epsilon)^{3/2}}{\sqrt{1-\epsilon}-\sqrt{\epsilon}\arcsin\sqrt{1-\epsilon}},
  \label{Om}
\end{equation}
(for $\epsilon <1$), see Appendix \ref{App:EqNodeLine}.
The average in Eq.~\eqref{eq:OptAnis} for the optimal variational anisotropy
factor can be reduced to
\begin{equation}
\left\langle \Omega^{2}\frac{v_{x}^{2}\!-\!\gamma_{o}^{2}v_{z}^{2}}{v_{x}^{2}\!+\!\gamma_{o}^{2}v_{z}^{2}}\right\rangle \!=\!1\!-\Omega_{0}^{2}\sqrt{\epsilon}\int\limits_{0}^{1}\frac{x^{2}dx}{\Gamma^{3/2}}\frac{2\gamma_{o}\epsilon x}{\sqrt{1\!-\!(1\!-\!\gamma_{o}^{2}\epsilon^{2})x^{2}}}\label{eq:gamoEqNodeInt}
\end{equation}
This integral can be taken analytically leading to a somewhat cumbersome result which is presented in Appendix \ref{App:EqNodeLine}. The angular average in Eq.~\eqref{eq:AnisVar} for the zero-temperature
$H_{c2}$ anisotropy can be reduced to the following integral
\begin{align}
&\left\langle \Omega^{2}\ln\left(\frac{v_{x}^{2}+\gamma_{o}^{2}v_{z}^{2}}{v_{x}^{2}+v_{y}^{2}}\right)\right\rangle\nonumber\\ &=2\Omega_{0}^{2}\sqrt{\epsilon}\int_{0}^{1}\frac{x^{2}dx}{\Gamma^{3/2}}\ln\frac{\gamma_{o}\epsilon x+\sqrt{1-\left(1-\gamma_{o}^{2}\epsilon^{2}\right)x^{2}}}{2\sqrt{1-x^{2}}},\label{eq:EqNodegamH}
\end{align}
which we compute numerically. The resulting dependence of $\gamma_{H}(0)$ vs $\epsilon$ is shown by the blue-dotted curve in Fig.\,\ref{EqLineNode}. We can see that it is smaller than $\gamma(T_c)$ meaning that it is reduced even stronger with respect to the Fermi surface anisotropy $1/\sqrt{\epsilon}$. Thus,  $\gamma_\lambda $ decreases on warming whereas  $\gamma_H $  is slightly increasing.

%

\subsection{ $\Omega=\Omega_0\sin \theta $, two polar point nodes }

\begin{figure}[tb]
\begin{center}
\includegraphics[width=8cm]{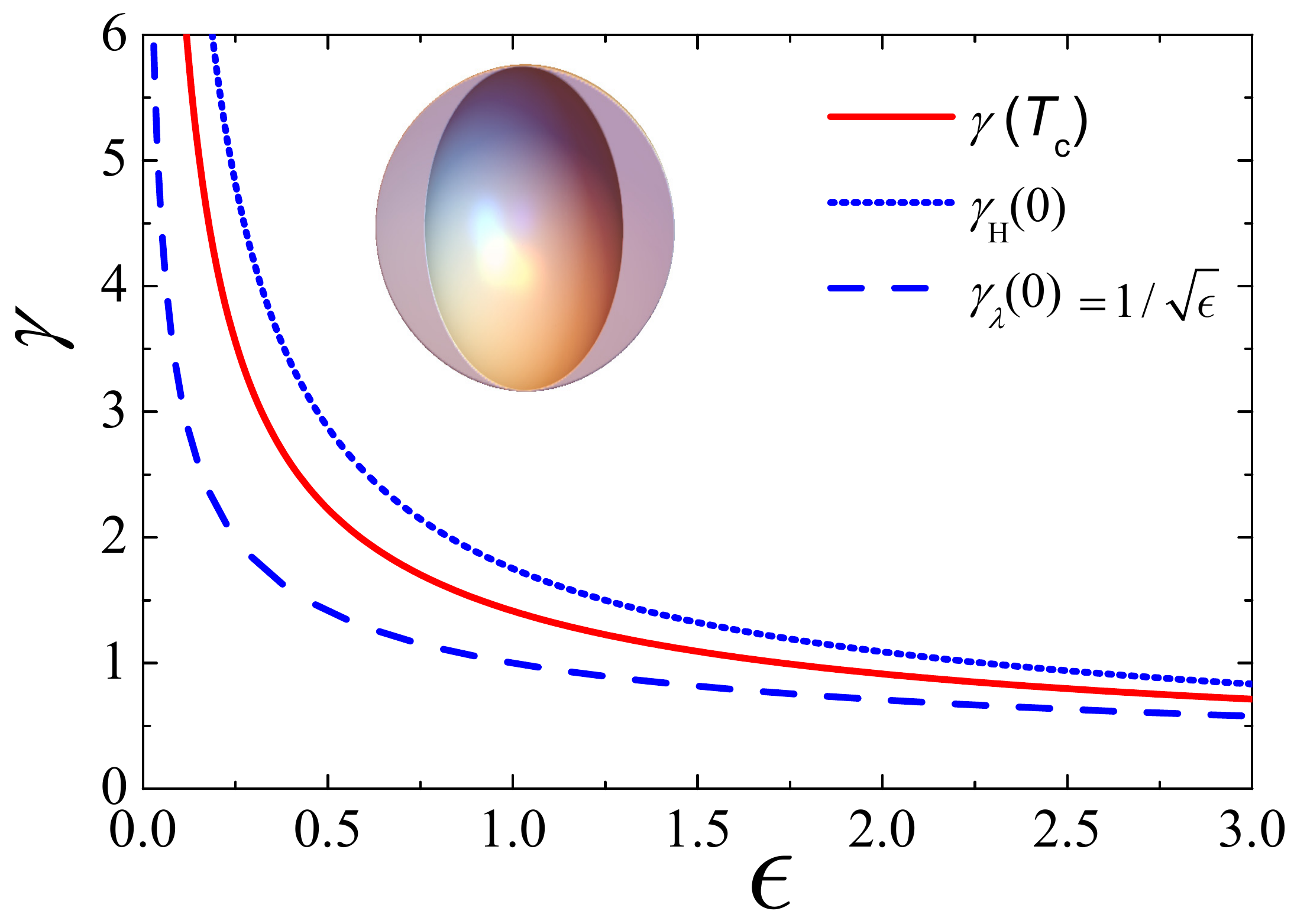} 
\caption{(Color online) $\gamma(\epsilon)$ for the order parameter with polar
		point nodes. The red, blue dashed and dotted-blue curve are $\gamma_(T_{c})$,
		 $\gamma_{\lambda}(0)=1/\sqrt{\epsilon}$, and  $\gamma_{H}(0)$, respectively }
	\label{FigPolarNodes}
\end{center}
\end{figure}

Two polar point nodes also may realize in the case of the p-wave order
parameter. Calculations in this case are similar to the previous one.
Similar to Eq.\ \eqref{gam_lamTc}, the anisotropy factor at $T_{c}$ is
\begin{equation}
\gamma^{2}(T_{c})=\frac{\int_{0}^{1}(1-x^{2})^{2}\Gamma^{-5/2}dx}{2\epsilon^{2}\int_{0}^{1}x^{2}(1-x^{2})\Gamma^{-5/2}dx}.
\label{gamTcPolar}
\end{equation}
The analytical result is given in Appendix \ref{App:PolarNodes}. The computed anisotropy factor
at $T_{c}$ is presented in Fig.\,\ref{FigPolarNodes} by the solid-red curve.
We can see that, in contrast to the case of equatorial node line, the order-parameter anisotropy
enlarges this factor in comparison with the Fermi-surface
anisotropy $\gamma_\lambda(0)=1/\sqrt{\epsilon}$ (blue dashed line). Consequently, $\gamma_{\lambda}$ increases on warming.

To compute $\gamma_{H}(0)$, we again have to evaluate averages in Eqs.~\eqref{eq:AnisVar}
and \eqref{eq:OptAnis}. The normalization constant has to be found
from the condition $\Omega_{0}^{2}\langle\sin^{2}\theta\rangle=1$.
As $\langle\sin^{2}\theta\rangle=1-\langle\cos^{2}\theta\rangle$,
we can use the result from Appendix \ref{App:PolarNodes} giving in the case $\epsilon<1$
\begin{equation}
\Omega_{0}^{2}=\frac{(1-\epsilon)^{3/2}}{\sqrt{\epsilon}\left(-\sqrt{\epsilon(1-\epsilon)}+\arcsin\sqrt{1-\epsilon}\right)}
.\label{OmPolar}
\end{equation}
The average in Eq.~\eqref{eq:OptAnis} for the optimal variational
anisotropy factor now becomes
\begin{equation}
\left\langle \!\Omega^{2}\frac{v_{x}^{2}\!-\!\gamma_{o}^{2}v_{z}^{2}}{v_{x}^{2}\!+\!\gamma_{o}^{2}v_{z}^{2}}\right\rangle
\!=\!1-\Omega_{0}^{2}\sqrt{\epsilon}\!\int\limits_{0}^{1}\!\!\frac{(1\!-\!x^{2})dx}{\Gamma^{3/2}}
\frac{2\gamma_{o}\epsilon x}
{\sqrt{1\!-\!(1\!-\!\gamma_{o}^{2}\epsilon^{2})x^{2}}}.
\label{eq:gamoPolar}
\end{equation}
We present the analytic result for this integral in Appendix
\ref{App:PolarNodes}. The angular average in Eq.~\eqref{eq:AnisVar}
for the $\gamma_{H}(0)$ can be reduced to the integral
\begin{align}
& \left\langle \Omega^{2}\ln\left(\frac{v_{x}^{2}+\gamma_{o}^{2}v_{z}^{2}}{v_{x}^{2}+v_{y}^{2}}\right)\right\rangle \nonumber \\
& =\!2\Omega_{0}^{2}\sqrt{\epsilon}\!\int\limits_{0}^{1}\!\frac{\left(1\!-\!x^{2}\right)dx}{\Gamma^{3/2}}\ln\frac{\gamma_{o}\epsilon x\!+\sqrt{1\!-\left(1\!-\!\gamma_{o}^{2}\epsilon^{2}\right)x^{2}}}{2\sqrt{1-x^{2}}},\label{eq:PolarNodegamH}
\end{align}
which we compute numerically. The calculated dependence of $\gamma_{H}(0)$
vs $\epsilon$ is shown in Fig.\,\ref{FigPolarNodes} by the blue-dotted
curve. We can see that it is larger than $\gamma(T_{c})$ meaning
that it is enlarged even stronger with respect to the Fermi-surface
anisotropy $1/\sqrt{\epsilon}$. Thus, $\gamma_{\lambda}$ increases
and $\gamma_{H}$ decreases on warming, opposite to the case of an equatorial node line.

\section{Two s-wave bands}

Many materials have multiple nonequivalent bands with different superconducting gaps. The most notable examples are magnesium diboride and iron-based superconductors.
The two-band model is the simplest model addressing this situation
allowing for qualitative understanding of multiple-band effects.
The iron-based superconductors  may have $\pm s$ symmetry meaning that the order parameter has opposite signs in different bands. The relative sign of order parameter, however, is irrelevant for the behavior of anisotropies in clean materials.


We consider two spheroidal Fermi surfaces with constant s-wave gaps. Therefore, each band is characterized by four parameters: the in-plane effective masses $m_{ab,\alpha}$, anisotropies $\epsilon_{\alpha}$, band depths $E_{\alpha}$ (distances between the Fermi level and the band's bottom or top), and gaps $\Delta_{\alpha}$.
Here and below $\alpha\!=\!1,2$ is the band's index.
Note that whether the band has an electron or hole character does not play any role in our consideration and $m_{ab,\alpha}$ notates here the absolute value of the effective mass.
Relative properties of the bands may be characterized by four ratio $r_m\!=\!m_{ab,2}/m_{ab,1}$, $r_\epsilon\!=\!\sqrt{\epsilon_{2}/\epsilon_{1}}$,  $r_E\!=\!E_{2}/E_{1}$, and $r_\Delta\!=\!\Delta_{2}^2/\Delta_{1}^2$.

This two-band model corresponds to the gap anisotropy given by
\begin{equation}
\Omega ({\bm k})= \Omega_{1,2}\,,\quad {\bm k}\in   F_{1,2} \,,
 \label{e50}
\end{equation}
where $F_1,F_2$ are two sheets of the Fermi surface and $\Omega_{1,2}$ are constants.  We denote
the densities of states (DOS) on the two parts as $N_{1,2}$,
\begin{equation}
N_\alpha= \frac{m_{ab,\alpha}^2v_{ab,\alpha}}{2\pi^2\hbar^3 \sqrt{\epsilon_\alpha}}  \,,\qquad \alpha=1,2\,.
\label{eq:bandDOS}
\end{equation}
Assuming   $X$ being constant at each sheet, we have:
\begin{equation}
\langle X \rangle = (X_1 N_1+X_2 N_2)/N(0) =  n_1X_1+n_2X_2,
\label{norm2}
\end{equation}
where $N(0)\!=\!N_1\!+\!N_2$,  $n_{1,2}= N_{1,2}/N(0)$ are  normalized DOS' with $n_1\!+\!n_2\!=\!1$ and their ratio $r_n\!=\!n_2/n_1$ is related to the above parameters $r_a$ as $r_n\!=\!r_{m}^{3/2}r_{E}^{1/2}r_{\epsilon}^{-1}$.  Since
the average over the full Fermi surface $\langle\Omega^2\rangle=1$, one has
\begin{equation}
n_1\Omega_1^2 +n_2\Omega_2^2=1\label{norm1}
\end{equation}
meaning that $\Omega_\alpha^2=\Delta_{\alpha}^2/(n_1\Delta_1^2 +n_2\Delta_2^2)$. The ratio of the 'superconducting band weights' $\zeta_\alpha \equiv n_\alpha \Omega_\alpha ^2$ is $\zeta_2/\zeta_1=r_nr_{\Delta}$.

Within this  model we obtain:
\begin{eqnarray}
 &&\gamma^2(T_c)= \frac{\langle\Omega ^2v_x^2\rangle}{\langle\Omega ^2v_z^2\rangle} = \frac{\zeta_1\left\langle  v_x^2 \right\rangle_1 +\zeta_2\left\langle  v_x^2 \right\rangle_2}{\zeta_1\left\langle  v_z^2 \right\rangle_1 +\zeta_2\left\langle  v_z^2 \right\rangle_2} ,\qquad \label{g-lam-Tc}\\
&&  \gamma_\lambda^2(0)=\frac{\langle v_x^2\rangle}{ \langle v_z^2\rangle} =\frac{n_1 \left\langle  v_x^2 \right\rangle_1 +n_2 \left\langle  v_x^2 \right\rangle_2}{n_1 \left\langle  v_z^2 \right\rangle_1 +n_2 \left\langle  v_z^2 \right\rangle_2} \,.
    \label{<gam_lam(0)>}
\end{eqnarray}
The  ratios of average squared velocities are  obtained using Appendix A and Eqs.\,(\ref{<vx2>}):
\begin{align*}
  \frac{\left\langle  v_x^2 \right\rangle_\alpha }{\left\langle  v_z^2 \right\rangle_\alpha } =&\frac{1}{\epsilon_\alpha},\
  r_v=\frac{\left\langle  v_x^2 \right\rangle_2 }{\left\langle  v_x^2 \right\rangle_1 } =\frac{v_{ab,2}^2}{v_{ab,1}^2}=\frac{r_{E}}{r_{m}},\ \\
  \frac{\left\langle  v_z^2 \right\rangle_2 }{\left\langle  v_z^2 \right\rangle_1 } =&\frac{v_{ab,2}^2\,\epsilon_2}{v_{ab,1}^2\,\epsilon_1}=r_vr_{\epsilon}^2.
\end{align*}
The results for the anisotropy factors look simpler when expressed via ratios $r_n$ and $r_v$ instead of $r_m$ and $r_E$.
We can express $\gamma(T_{c})$ and $\gamma_{\lambda}(0)$ in terms of the introduced ratios as
\begin{align}
\gamma^{2}(T_{c})\epsilon_{1}=&
\frac{1+r_{n}r_{v}r_{\Delta}}{1+r_{n}r_{v}r_{\Delta}r_{\epsilon}^2}, \label{gam-Tc-Ratios}\\
\gamma_{\lambda}^{2}(0)\epsilon_{1}=&\frac{1+r_{n}r_{v}}{1+r_{n}r_{v}r_{\epsilon}^2}\,.
 \label{gam_lam0Ratio}
\end{align}

The variational estimate for the low-temperature $H_{c2}$ anisotropy $\gamma_H(0)$ is described in Appendix \ref{App:TwoBands} and leads to the following result
\begin{equation}
\gamma_{H}(0)\sqrt{\epsilon_{1}}\approx\kappa_{o1}\left(\frac{1\!+\!\kappa_{o1}}{2}\right)^{-2\zeta_{1}}
\!\!\left(\frac{1\!+\!r_{\epsilon}\kappa_{o1}}{2}\right)^{-2\zeta_{2}}
\label{eq:gamH0TwoBands}
\end{equation}
with $\zeta_{1}=1-\zeta_{2}=(1+r_{n}r_{\Delta})^{-1}$,
\begin{equation}
\kappa_{o1}\!\equiv\!\gamma_{o}\sqrt{\epsilon_{1}}\!=\!\frac{\zeta_{-}}{2}\left(1\!-\frac{1}{r_{\epsilon}}\right)\!+\!\sqrt{\frac{\zeta_{-}^{2}}{4}\left(1\!+\frac{1}{r_{\epsilon}}\right)^{2}\!+\frac{4\zeta_{1}\zeta_{2}}{r_{\epsilon}}}
\label{eq:gvarTwoBands}
\end{equation}
and $\zeta_{-}=\zeta_{1}-\zeta_{2}$. Comparing these equation with Eq.\ \eqref{gam-Tc-Ratios}, we see that the ratio $\gamma_{H}(0)/\gamma(T_{c})$ is determined by only three parameters, $r_\epsilon$, $r_v$, and the product  $r_n r_\Delta$ corresponding to the ratio of the condensation energies, $r_n r_\Delta=N_2\Delta_2^2/N_1\Delta_1^2$.

Equations \eqref{gam-Tc-Ratios}, \eqref{gam_lam0Ratio}, \eqref{eq:gamH0TwoBands}, and  \eqref{eq:gvarTwoBands} give general results for the anisotropy factors of two-band s-wave superconductors in terms of the band-parameter ratios. As the band numbering is arbitrary, all anisotropies are invariant with respect to the substitutions $\epsilon_{1}\leftrightarrow\epsilon_{2}$ and $r_a\rightarrow 1/r_a$ for all ratios. For definiteness, we will assume that the second band has higher anisotropy, i.e., $r_\epsilon<1$. We observe that, as expected, in the case of identical ellipticities $\epsilon_1=\epsilon_2=\epsilon$ all anisotropies are temperature independent and equal to $1/\sqrt{\epsilon}$ independently of other ratios.
As function of the DOS' ratio $r_n$, all three anisotropy factors interpolate in between $\gamma_1=1/\sqrt{\epsilon_1}$ for $r_n\rightarrow 0$ and $\gamma_2=1/\sqrt{\epsilon_2}$ for $r_n\rightarrow \infty$.
At intermediate $r_n$, however,  we can observe that the bands contribute  to different anisotropies with different weights, and one may have many situations.

The simplicity of this model notwithstanding, when applied, e.g.,  to the problem of $T$-dependence of anisotropies in MgB$_2$ \cite{gam-lam0}, it reproduces well the observed behavior.
The unique feature of MgB$_2$ is that $\gamma_\lambda$ and $\gamma_H$ have opposite temperature dependences: $\gamma_\lambda$ increases with warming from $\sim$$1$ to $\sim$$2$, while $\gamma_H$ drops from $\sim$$5-6$ to $\sim$$2$ \cite{AngstPRL02,LyardPRB02,RydhPRB04,Carrington}.
This behavior is consistent with multiband superconducting properties of this material. It has two groups of bands, three-dimensional  $\pi$-bands with smaller gap and quasi-two-dimensional  $\sigma$-bands with larger gap.  The band parameters are \cite{gam-lam0} $\epsilon_{1}\approx 1.3$ ($\pi$-band), $\epsilon_{2}\approx 0.022$ ($\sigma$-band), $r_n\approx 0.8$, $r_v\approx 0.7$, and $r_\Delta\approx 16$. Then Eqs.\ \eqref{gam-Tc-Ratios}, \eqref{gam_lam0Ratio}, and \eqref{eq:gamH0TwoBands} give $\gamma(T_c)\approx 2.6$,  $\gamma_\lambda(0)\approx 1.1$, and $\gamma_H(0)\approx 5.5$, which is roughly consistent with experiment.

\begin{figure}[tb]
	\begin{center}
		\includegraphics[width=8cm] {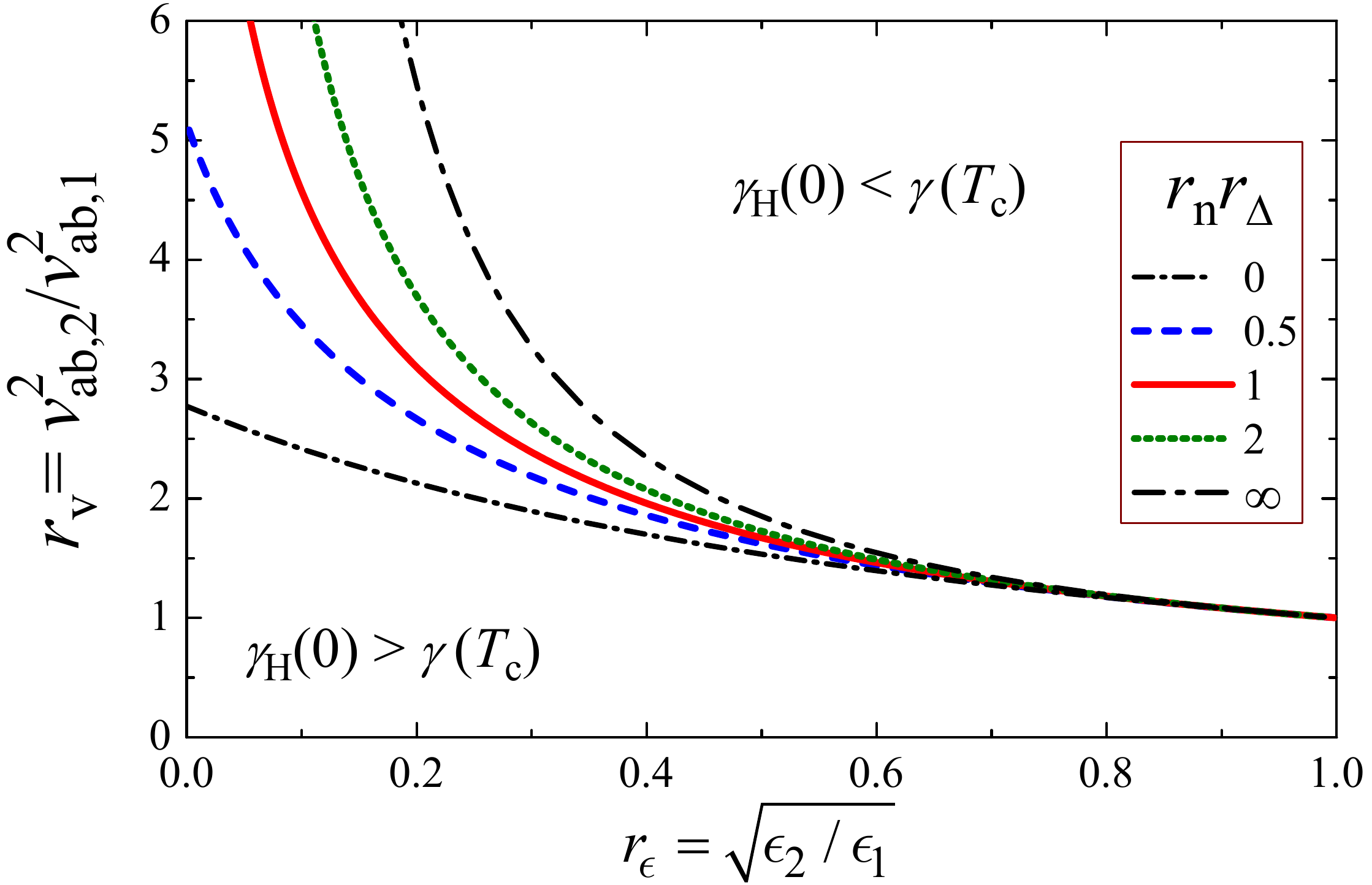}
		\caption{ (Color online) The dependence of the Fermi-velocity ratio separating the two regimes of $\gamma_H(T)$ on the anisotropy ratio $r_\epsilon$ for three values of the condensation-energy ratio $r_nr_\Delta$, $0.5$, $1$, and $2$. The dash-dotted lines show the limiting values defined by Eqs.\ \eqref{eq:rvmin} and \eqref{eq:rvmax}.
		}
		\label{fig:rv-re}
	\end{center}
\end{figure}

The most interesting question is what band properties determine the direction of temperature dependences of the anisotropy factors. In the case of $\gamma_\lambda$ this question has a straightforward and simple answer.
From Eqs.\ \eqref{gam-Tc-Ratios} and \eqref{gam_lam0Ratio}, we derive a relation
\begin{equation}
\frac{\gamma_{\lambda}^{2}(0)}{\gamma^{2}(T_{c})}-1=\frac{r_{n}r_{v}\left(r_{\epsilon}^{2}-1\right)\left(r_{\Delta}-1\right)}{\left(1+r_{n}r_{v}r_{\epsilon}^{2}\right)\left(1+r_{n}r_{v}r_{\Delta}\right)},
\label{eq:gLamRatio2Bands}
\end{equation}
which shows that the direction of the $\gamma_{\lambda}$ temperature dependence is determined only by relations between the bands anisotropies and gaps. In particular, $\gamma_{\lambda}(T)$ increases on warming if the band with higher anisotropy has also larger gap (like in MgB$_2$), and decreases otherwise.

The case of $\gamma_H$ is more involved. The simple criterion can be obtained only in the case of small difference between the ellipticities $|\epsilon_{1}-\epsilon_{2}|\ll \epsilon_{1}$. In this case expansion with respect to small parameter $r_\epsilon-1$ gives
\begin{equation}
\frac{\gamma_{H}(0)}{\gamma(T_{c})}-1\approx\frac{r_{n}r_{\Delta}\left(r_{v}-1\right)\left(r_{\epsilon}-1\right)}{\left(1+r_{n}r_{\Delta}\right)\left(1+r_{n}r_{v}r_{\Delta}\right)}
\label{eq:gHRatio2BandsExp}
\end{equation}
indicating that $\gamma_H(T)$ increases on warming if the band with higher anisotropy has larger Fermi velocity. When the band anisotropies are not close, there is no simple criterion determining the direction of the $\gamma_H$ temperature dependence. In the case $r_\epsilon<1$,  $\gamma_H(T)$ decreases with temperature at very small Fermi-velocity ratios $r_v$ and vice versa. The value of $r_v$ separating the two regimes depends on two parameters, $r_\epsilon$ and $r_nr_\Delta$.  Analytical results for this quantity can be derived in the limiting cases  $r_nr_\Delta \rightarrow 0$ and $\infty$.
\begin{align}
r_{v,\mathrm{min}}=&\frac{4}{1-r_{\epsilon}^{2}}\ln\left(\frac{2}{1+r_{\epsilon}}\right), \text{ for }r_nr_\Delta \rightarrow 0\label{eq:rvmin}\\
r_{v,\mathrm{max}}=&\frac{r_{\epsilon}^{-2}-1}{4\ln\left[\left(r_{\epsilon}^{-1}\!+\!1\right)\!/2\right]},
\text{ for }r_nr_\Delta \rightarrow \infty\label{eq:rvmax}
\end{align}
Figure \ref{fig:rv-re} shows these limiting velocity ratios together with the numerically computed dependences   $r_v(r_\epsilon)$ for three intermediate values of the condensation-energy ratio $r_nr_\Delta$, $0.5$, $1$, and $2$. We can see that the separating $r_v$ increases with decreasing $r_\epsilon$ and also grows with increasing $r_nr_\Delta$. The sensitivity to the latter parameter increases with decreasing $r_\epsilon$.
In particular, MgB$_2$ is located in the lower left corner of this plot.

\begin{figure}[tb]
	\begin{center}
		\includegraphics[width=8cm] {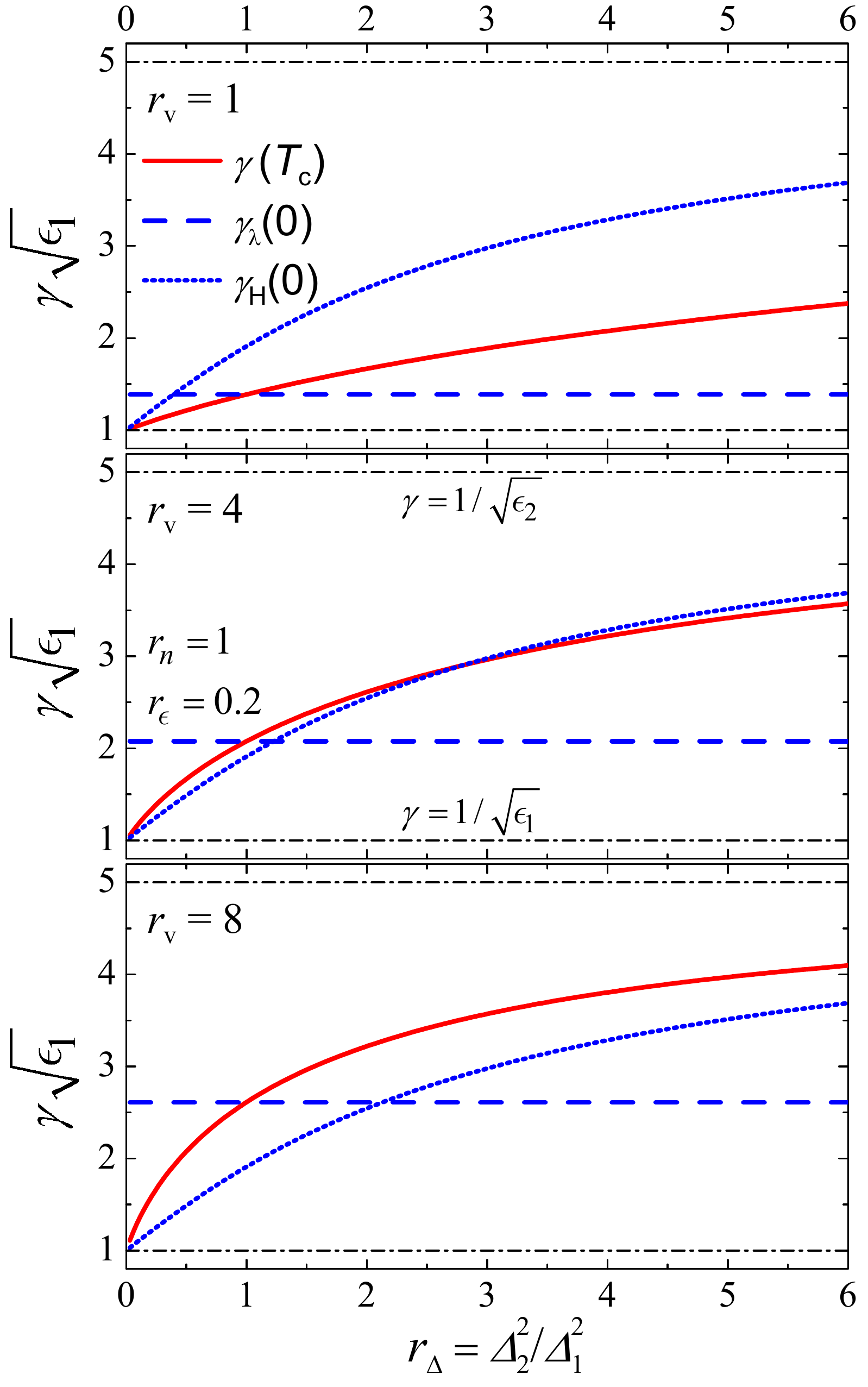}
		\caption{ (Color online) The dependences of the anisotropy factors on the relative strength of superconductivity in two bands for three representative cases showing qualitatively different behavior. All plots are made for $r_n\!=\!1$ and $r_\epsilon\!=\!0.2$.  In the upper plot $r_v\!=\!1$ and in this case $\gamma_H$ decreases with temperature in the whole range of $r_\Delta$. The lower plot is made for much higher velocity ratio $r_v\!=\!8$ and in this case $\gamma_H$ always increases with temperature.
		The middle plot is made for the intermediate velocity ratio $r_v\!=\!4$, for which $\gamma_H(0)$ and $\gamma(T_c)$ switch with increasing $r_\Delta$.
		 The dash-dotted lines show the anisotropy factors of the two bands.
		}
		\label{fig:g-rD}
	\end{center}
\end{figure}

To illustrate typical behaviors of the anisotropy factors, we present in Fig.\ \ref{fig:g-rD} their dependences on the ratio $r_\Delta$ characterizing a relative strength of superconductivity in two bands. The plots are made for $r_n\!=\!1$, $r_\epsilon\!=\!0.2$ and three values of the Fermi velocity ratio $r_v$, $1$, $4$, and $8$. In the first case $r_v$ is below $r_{v,\mathrm{min}}$ in Eq.\ \eqref{eq:rvmin} meaning that
$\gamma_H(0)$ exceeds $\gamma(T_c)$ in the whole range of $r_\Delta$ and $\gamma_H$ always decreases on warming. In the range $r_\Delta>1 $ such situation qualitatively corresponds to magnesium diboride.
The last value $r_v\!=\!8$ exceeds $r_{v,\mathrm{max}}$ given by Eq.\ \eqref{eq:rvmax} and the behavior is opposite, $\gamma_H$ always increases with temperature.
The intermediate value $r_v\!=\!4$ is in between $r_{v,\mathrm{min}}$ and $r_{v,\mathrm{max}}$. In this case
the behavior of $\gamma_H$ switches with increasing $r_\Delta$: $\gamma_H(0)$ is smaller than $\gamma(T_c)$ for small $r_\Delta$ and exceeds  $\gamma(T_c)$ for  $r_\Delta > 2.8$.
While $\gamma_\lambda(0)$ does not depend on $r_\Delta$, both $\gamma_H(0)$ and $\gamma (T_c)$ increase with  $r_\Delta$ interpolating between $1/\sqrt{\epsilon_1}$ for $r_\Delta\rightarrow 0$ and $1/\sqrt{\epsilon_2}$  for $r_\Delta\rightarrow \infty$.  Consequently, the latter two anisotropies always cross $\gamma_\lambda(0)$.
In particular, as mentioned above, $\gamma_\lambda$ decreases on warming at $r_\Delta<1$ and increases with temperature at $r_\Delta>1$. Therefore, depending on the band parameters, all possible six relations between the three anisotropy factors may be realized.

\section{Summary}

It is shown that the anisotropy parameters, $\gamma_H $ for the upper critical field  and $\gamma_\lambda$ for the penetration depth,  in general, depend on temperature even in one-band materials with anisotropic order parameters. The temperature behavior of $\gamma$'s depends, in particular, on the order parameter nodes and their distributions.

We provide four examples of gap anisotropies. In the simplest reference case of isotropic s-wave, $\gamma_\lambda=\gamma_H =1/\sqrt{\epsilon}$ and is $T$ independent. For d-wave symmetry with coinciding polar axis of the order parameter and of Fermi spheroid, $\gamma_\lambda$ is the same as for s-wave and is $T$ independent.  $\gamma_H $ is weakly changing  on warming, with the sign of this change depending on what direction of in-plane field is chosen for determination of the anisotropy parameter, see Fig.\,\ref{f1new}.
If $\Delta$ has a line node on the equator of Fermi spheroid,  $\gamma_\lambda(T)$ decreases on warming  whereas $\gamma_H(T)$ increases, see Fig.\,\ref{EqLineNode}.
Point nodes at spheroid poles affect anisotropies in opposite way as demonstrated in Fig.\,\ref{FigPolarNodes}. In general, the former/latter behavior is realized in the cases when the gap at the equator is smaller/larger than the gap at the poles.

In the case of two spheroidal Fermi surfaces, we investigated in detail the conditions determining directions of temperature dependences of the anisotropy factors. We found that $\gamma_\lambda$ increases with temperature only if the band with higher anisotropy has larger gap independently on relations between other band parameters.  The behavior of $\gamma_H$ is more complicated. In general, $\gamma_H$ increases with temperature if the Fermi velocity of the more anisotropic band sufficiently exceeds the Fermi velocity of the less anisotropic band. The Fermi-velocity ratio separating the two regimes depends in a nontrivial way on the ratios of band anisotropies and condensation energies, as illustrated in Fig.\ \ref{fig:rv-re}.  In general, all six possible relations between three key anisotropy factors, $\gamma(T_c)$, $\gamma_\lambda(0)$, and $\gamma_H(0)$, may be realized for different relations between band parameters, see Fig.\ \ref{fig:g-rD}.

Our results for Fermi ellipsoids are based on theory \cite{ROPP}, which is a generalization of Helfand-Werthamer work \cite{HW} for the isotropic case. One of the features of this approach is that solutions of the  linear equation $- \xi^2_{ik}  \Pi_i\Pi_k \Psi =\Psi$ at $H_{c2}$, belonging to the lowest Landau level, satisfy also the self-consistency equation for superconductivity. For general Fermi surfaces this is not the case and the exact solution can be obtained using expansion over full set of the Landau-level wave functions, see e.g. Refs.\ \cite{RieckPhysB90,Kita}.  In this paper,  we employ a much simpler approximate variational approach using the lowest-Landau wave function as a trial. This approach leads to reasonable results suitable for  qualitative interpretations of data on anisotropy parameters.

Also, it is worth keeping in mind that   we estimate the anisotropy of {\it orbital} $H_{c2}$  and disregard  possibility of Pauli limiting effects.
The decreasing $H_{c2}$ anisotropy with decreasing temperature, like, e.g., in iron pnictides \cite{Yuan2009,Khim2011,Meier2016}, is usually considered as an indication of strong paramagnetic effect. While this interpretation in many cases is correct, we point out that such a behavior may also be realized in purely orbital case, as illustrated in Figs.\ \ref{EqLineNode} and \ref{fig:g-rD}.
Even though modeling  many-band systems with two Fermi ellipsoids is far from being realistic, this simple approach still provides some straightforward inroads to a complicated interplay of anisotropies $\gamma_\lambda(T)$ and $\gamma_H(T)$.

\begin{acknowledgments}
This work was supported by the U.S. Department of Energy (DOE), Office of Science, Basic Energy Sciences, Materials Science and Engineering Division. The research of VK and RP was performed at Ames Laboratory, which is operated for the U.S. DOE by Iowa State University under contract \# DE-AC02-07CH11358.   The work of AK in Argonne was supported by the U.S. Department of Energy, Office of Science, Basic Energy Sciences,  Materials Sciences and Engineering Division.
\end{acknowledgments}

\appendix

 \section{Fermi spheroid}

The Fermi surface as an ellipsoid of rotation is an interesting example in its own right and as a model  for calculating $H_{c2}$ and $\lambda$ in uniaxial materials. Since both of these quantities are derived employing integrals over the full Fermi surface, they are weakly sensitive to fine details of Fermi surfaces.

Although straightforward, the averaging over the Fermi spheroid of the main text   should be done with care. Moreover, there are   examples in literature where these averages were done incorrectly   \cite{MMK,ROPP}. We reproduce here this procedure to correct the error and to prevent it in future.

Consider an  uniaxial superconductor with the electronic spectrum
\begin{equation}
E(\bm k)=\hbar^2\left(\frac{k_x^2+k_y^2}{2m_{ab}}+\frac{k_z^2}{2m_c}\right)\,,
\end{equation}
so that the Fermi surface is an ellipsoid of rotation with $z$ as the symmetry axis.
In spherical coordinates $(k,\theta,\phi)$
\begin{equation}
E =\frac{\hbar^2k^2}{2m_{ab}}\left(
\sin^2\theta+\frac{m_{ab}}{m_c}\cos^2\theta
\right)=\frac{\hbar^2k^2}{2m_{ab}}\Gamma(\theta),
\end{equation}
so that
\begin{equation}
k_F^2(\theta)=\frac{ 2m_{ab} E_F }{\hbar^2 \Gamma(\theta)} \,.
\label{kF}
\end{equation}
The Fermi velocity is   $\bm v(\bm k)=\bm\nabla_{\bm k}E(\bm k)/\hbar$,
with the derivatives taken at $\bm k=\bm k_F$:
\begin{eqnarray}
v_x&=&\frac{v_{ab} \sin\theta\cos\phi}{\sqrt{\Gamma(\theta)}}, \,\,\,
v_y=\frac{v_{ab}\sin\theta\sin\phi}{\sqrt{\Gamma(\theta)}},\,\\
 v_z&=&\epsilon \frac{v_{ab}\cos\theta}{\sqrt{\Gamma(\theta)}},\quad
 \label{velocities}
 \epsilon=\frac{m_{ab}}{m_c}\,,   \quad v_{ab}= \sqrt{\frac{2E_F}{m_{ab}}}\,.\qquad
\end{eqnarray}
The  value of the Fermi
velocity,  $v =(v_x^2+v_y^2+v_z^2)^{1/2}$, is
\begin{equation}
v= v_{ab}\sqrt{\frac{
 \sin^2\theta+\epsilon^2 \cos^2\theta
}{\sin^2\theta+\epsilon\cos^2\theta}}  \,.
\end{equation}

The density of states $N(0)$ is defined as an integral over the Fermi surface:
\begin{equation}
N(0)=\int \frac{\hbar^2d^2\bm k_F}{(2\pi\hbar)^3v} \,.
 \label{Koshelev}
\end{equation}
 An area element of  the spheroid surface is
\begin{equation}
d^2\bm k_F=k_F(\theta)\sqrt{k_F^2(\theta) +\left(\frac {dk_F}{d\theta}\right)^2}
\sin \theta\,d\theta\,d\varphi
 \label{Kosh}
\end{equation}
and  after simple algebra one obtains:
\begin{equation}
\frac{d^2\bm k_F}{v}= \frac {k_{F,ab}^2}{v_{ab}} \frac{\sin \theta\,d\theta\,d\varphi}{\Gamma^{3/2}}
 \label{a9}
\end{equation}
where $k_{F,ab}^2=2m_{ab}E_F/\hbar^2$. This gives
\begin{equation}
N(0)= \frac{m_{ab}^2v_{ab}}{2\pi^2\hbar^3 \sqrt{\epsilon}}  \,.
\label{DOS}
\end{equation}

 The normalized local density of states within solid angle $\sin \theta\,d\theta\,d\varphi$ is
\begin{equation}
\frac{\hbar^2d^2\bm k_F}{(2\pi\hbar)^3vN(0)}=\frac{\sqrt{\epsilon}}{4\pi}
  \frac{\sin \theta\,d\theta\,d\varphi}{\Gamma^{3/2}} \,;
 \label{localDOS}
\end{equation}
 Eq.\,(\ref{a9}) has been used here. Thus, the Fermi surface average of a function $A( \theta,\phi)$   is:
 \begin{eqnarray}
\left\langle A( \theta,\phi) \right\rangle =\frac{\sqrt{\epsilon}}{4\pi }
\int A( \theta,\varphi) \frac{\sin \theta\,d\theta\,d\varphi}{\Gamma^{3/2}(\theta)} \,.
 \label{averageApp}
\end{eqnarray}


\section{Variational estimate of the upper critical field}\label{App-var}

Equation for the upper critical field \eqref{selfcons3} has exact
solution only in few special cases. In general situation the exact
numerical solution may be obtained, for example, by expansion over
a complete set of Landau-level wave functions \cite{RieckPhysB90,MMK}.
An approximate solution giving in many cases a reasonable accuracy
may be obtained using the variational approach \cite{DahmPRL03}.
The equation \eqref{selfcons3} corresponds to the following variational
problem
\begin{widetext}
\begin{equation}
\ln t=\max\frac{\int d\boldsymbol{r}\psi_{t}\left(\boldsymbol{r}\right)\int_{0}^{\infty}d\rho\,\ln\tanh\frac{\pi T\rho}{\hbar}\,\left\langle \Omega^{2}{\bf v}\cdot\bm{\Pi}e^{-\rho{\bf v}\cdot\bm{\Pi}}\right\rangle \psi_{t}\left(\boldsymbol{r}\right)}{\int d\boldsymbol{r}\psi_{t}^{2}\left(\boldsymbol{r}\right)},\label{eq:VarEq}
\end{equation}
\end{widetext}
where the maximum has to be found over all possible trial functions
$\psi_{t}\left(\boldsymbol{r}\right)$. Consider, for definiteness,
the magnetic field along the $y$ axis. The simplest and most natural
choice for the trial function is the lowest Landau-level solution
of anisotropic equation for particle with charge $2e$ in magnetic
field
\begin{equation}
-\left(\Pi_{x}^{2}+\gamma_{t}^{-2}\Pi_{z}^{2}\right)\psi_{t0}=\frac{2\pi H}{\Phi_{0}\gamma_{t}}\psi_{t0}\label{eq:trialLL}
\end{equation}
with $\Pi_{x}\!=\!\partial_{x}$, $\Pi_{z}\!=\!\partial_{z}\!+\!2\pi iHx/\Phi_{0}$,
where the anisotropy factor $\gamma_{t}$ is the variational parameter.
We will assume that $\psi_{t0}\left(\boldsymbol{r}\right)$ is normalized,
$\int d\boldsymbol{r}\psi_{t0}^{2}\left(\boldsymbol{r}\right)=1$.

The integral in Eq. \eqref{eq:VarEq} is determined by the matrix
element $\left\langle \psi_{t0}|{\bf v}\cdot\bm{\Pi}e^{-\rho{\bf v}\cdot\bm{\Pi}}|\psi_{t0}\right\rangle _{r}$,
where we use notation $\left\langle \psi_{t0}\hat{|U}|\psi_{t0}\right\rangle _{r}=\smallint d\boldsymbol{r}\psi_{t0}(\boldsymbol{r})\hat{U}\psi_{t0}(\boldsymbol{r})$.
To evaluate this matrix element, we introduce the operators $\mathcal{P}_{\pm}=\Pi_{x}\pm i\gamma_{t}^{-1}\Pi_{z}$
and represent the product ${\bf v}\cdot\bm{\Pi}$ as
\begin{align*}
{\bf v}\cdot\bm{\Pi} & =\frac{v_{-}\mathcal{P}_{+}+v_{+}\mathcal{P}_{-}}{2},
\end{align*}
where $v_{\pm}=v_{x}\pm i\gamma_{t}v_{z}$.
Note that $\mathcal{P}_{-}\psi_{t0}=0$. Using the relations $e^{P+Q}\,=e^{P}e^{Q}e^{[Q,P]/2}$,
$\left[\mathcal{P}_{-},\mathcal{P}_{+}\right]=-4\pi H/\gamma_{t}\Phi_{0}$,
and $v_{+}v_{-}=v_{x}^{2}+\gamma_{t}^{2}v_{z}^{2}$, we transform
\begin{align*}
e^{-\rho{\bf v}\cdot\bm{\Pi}} & =\exp\left(-\frac{\rho v_{-}\mathcal{P}_{+}}{2}\right)\exp\left(-\frac{\rho v_{+}\mathcal{P}_{-}}{2}\right)\\
\times&\exp\left(-\frac{\pi}{2}\left(v_{x}^{2}+\gamma_{t}^{2}v_{z}^{2}\right)\frac{H\rho^{2}}{\gamma_{t}\Phi_{0}}\right).
\end{align*}
This relation allows us to evaluate the matrix element as
\begin{widetext}
\[
\left\langle \psi_{t0}|{\bf v}\cdot\bm{\Pi}e^{-\rho{\bf v}\cdot\bm{\Pi}}|\psi_{t0}\right\rangle _{r}  =\left\langle \psi_{t0}|\frac{v_{-}\mathcal{P}_{+}+v_{+}\mathcal{P}_{-}}{2}\exp\left(-\frac{\rho v_{-}\mathcal{P}_{+}}{2}\right)\exp\left(-\frac{\rho v_{+}\mathcal{P}_{-}}{2}\right)|\psi_{t0}\right\rangle _{r}
 =\left(v_{x}^{2}+\gamma_{t}^{2}v_{z}^{2}\right)\frac{\pi H\rho}{\Phi_{0}\gamma_{t}},
\]	
leading to the variational equation for the upper critical field,
\begin{equation}
\ln t=\max_{\gamma_{t}}\int_{0}^{\infty}d\rho\,\ln\tanh\frac{\pi T\rho}{\hbar}\,\left\langle \Omega^{2}\exp\left[-\frac{\pi}{2}\left(v_{x}^{2}+\gamma_{t}^{2}v_{z}^{2}\right)\frac{H\rho^{2}}{\Phi_{0}\gamma_{t}}\right]\left(v_{x}^{2}+\gamma_{t}^{2}v_{z}^{2}\right)\frac{\pi H\rho}{\Phi_{0}\gamma_{t}}\right\rangle \,.\label{eq:Hc2Appr-1}
\end{equation}
In the zero-temperature limit this equation becomes
\begin{equation}
\max_{\gamma_{t}}\int_{0}^{\infty}d\rho\,\ln\frac{\pi T_{c}\rho}{\hbar}\,\left\langle \Omega^{2}\exp\left[-\frac{\pi}{2}\left(v_{x}^{2}+\gamma_{t}^{2}v_{z}^{2}\right)\frac{H\rho^{2}}{\Phi_{0}\gamma_{t}}\right]\left(v_{x}^{2}+\gamma_{t}^{2}v_{z}^{2}\right)\frac{\pi H\rho}{\Phi_{0}\gamma_{t}}\right\rangle=0 \,.\label{eq:Hc2Appr-1-1}
\end{equation}
\end{widetext}
As $\int_{0}^{\infty}d\rho\ln\left(\frac{\rho}{2}\right)2a\rho\exp\left[-a\rho^{2}\right]=\frac{1}{2}\left(-\boldsymbol{C}-\ln4a\right)$
with $\boldsymbol{C}\approx0.5772$, we obtain
\[
\min_{\gamma_{t}}\left\langle \Omega^{2}\ln\left[\left(v_{x}^{2}+\gamma_{t}^{2}v_{z}^{2}\right)\frac{H\hbar^{2}e^{\boldsymbol{C}}}{2\pi T_{c}^{2}\Phi_{0}\gamma_{t}}\right]\right\rangle =0.
\]
Introducing typical velocity scale $u_{x}^{2}=\left\langle \Omega^{2}v_{x}^{2}\right\rangle $,
we finally arrive at
\begin{align}
&H_{c2,y}(0)\approx\frac{2\pi T_{c}^{2}\Phi_{0}}{u_{x}^{2}\hbar^{2}e^{\boldsymbol{C}}}\nonumber\\
&\times\max_{\gamma_{t}}\left\{ \gamma_{t}\exp\left[-\left\langle \Omega^{2}\ln\left(\frac{v_{x}^{2}+\gamma_{t}^{2}v_{z}^{2}}{u_{x}^{2}}\right)\right\rangle \right]\right\} .\label{eq:Hc2yVar}
\end{align}
The optimal anisotropy factor $\gamma_{o}$ is determined by equation
\[
\left\langle \Omega^{2}\frac{d}{d\gamma_{t}}\ln\left[\frac{v_{x}^{2}+\gamma_{t}^{2}v_{z}^{2}}{\gamma_{t}}\right]\right\rangle =0,
\]
giving
\begin{equation}
\left\langle \Omega^{2}\frac{v_{x}^{2}-\gamma_{o}^{2}v_{z}^{2}}{v_{x}^{2}+\gamma_{o}^{2}v_{z}^{2}}\right\rangle =0.\label{eq:OptAnisApp}
\end{equation}
It is straightforward to demonstrate that in the case of single spheroidal
Fermi surface and isotropic order parameter, $\gamma_{o}=1/\sqrt{\epsilon}$
and Eq. \eqref{eq:Hc2yVar} gives the exact $y$-axis upper critical
field.

In this paper we only consider crystals isotropic within the $xy$
plane. In this case the optimal anisotropy factor for field along
the $z$ axis is obviously equal to one and
\begin{equation}
H_{c2,z}(0)\approx\frac{2\pi T_{c}^{2}\Phi_{0}}{u_{x}^{2}\hbar^{2}e^{\boldsymbol{C}}}\exp\left[-\left\langle \Omega^{2}\ln\left(\frac{v_{x}^{2}+v_{y}^{2}}{u_{x}^{2}}\right)\right\rangle \right].\label{eq:Hc2z}
\end{equation}
From Eqs. \eqref{eq:Hc2yVar} and \eqref{eq:Hc2z}, we obtain the
variational estimate for the low-temperature anisotropy factor
\begin{equation}
\gamma_{H}(0)=\gamma_{o}\exp\left[-\left\langle \Omega^{2}\ln\left(\frac{v_{x}^{2}+\gamma_{o}^{2}v_{z}^{2}}{v_{x}^{2}+v_{y}^{2}}\right)\right\rangle \right],\label{eq:AnisVarApp}
\end{equation}
where $\gamma_{o}$ has to be evaluated from Eq. \eqref{eq:OptAnisApp}.

\subsection{Singe-band cases}

\subsubsection{D-wave order parameter}\label{App-dwave}

In this subsection we summarize the calculations for the case of d-wave
order parameter for which the weight function has the form $\Omega(\varphi)=\sqrt{2}\cos2(\varphi-\varphi_{0})$
with $\varphi_{0}$ being the angle between the in-plane component
of the magnetic field and the direction of maximum order parameter. In this case evaluation
of the integral in Eq. \eqref{eq:OptAnis} gives
\begin{widetext}
	\begin{align*}
	\left\langle \Omega^{2}\frac{v_{x}^{2}\!-\!\gamma_{o}^{2}v_{z}^{2}}{v_{x}^{2}\!+\!\gamma_{o}^{2}v_{z}^{2}}\right\rangle  & \!=\frac{\sqrt{\epsilon}}{4\pi}\int\limits _{0}^{\pi}\frac{\sin\theta d\theta}{\left(\sin^{2}\theta\!+\!\epsilon\cos^{2}\theta\right)^{3/2}}\int\limits _{-\pi}^{\pi}d\varphi2\cos^{2}2(\varphi\!-\!\varphi_{0})\frac{\sin^{2}\theta\cos^{2}\varphi\!-\!\gamma_{o}^{2}\epsilon^{2}\cos^{2}\theta}{\sin^{2}\theta\cos^{2}\varphi\!+\!\gamma_{o}^{2}\epsilon^{2}\cos^{2}\theta}\\
	& =\!1\!-\!4\cos\left(4\varphi_{0}\right)\kappa_{o}^{2}\left[2\left(3\kappa_{o}^{2}\!-\!1\right)\ln\left(1\!+\!\frac{1}{\kappa_{o}}\right)-3(2\kappa_{o}\!-\!1)\right]-\left(1\!+\!\cos\left(4\varphi_{0}\right)\right)\frac{2\kappa_{o}}{1\!+\!\kappa_{o}}
	\end{align*}
	with $\kappa_{o}^{2}=\gamma_{o}^{2}\epsilon$. We find that numerical
	solution of Eq. \eqref{eq:OptAnis} is very close to the simple dependence
	$\kappa_{o}(\varphi_{0})\approx1-0.181\cos\left(4\varphi_{0}\right)$.
	Note that the result $\kappa_o(\pi/8)=1$ is exact.
	
	The average in Eq. \eqref{eq:AnisVar} can be evaluated as
	\[
	\left\langle \Omega^{2}\ln\left(\frac{v_{x}^{2}\!+\!\gamma_{o}^{2}v_{z}^{2}}{v_{x}^{2}+v_{y}^{2}}\right)\right\rangle =\!2\ln\left(\frac{1\!+\!\kappa_{o}}{2}\right)\!+\!\cos\left(4\varphi_{0}\right)\left[\frac{1}{2}-(2\kappa_{o}\!-\!1)\left(3\kappa_{o}^{2}\!-\!1\right)\!-\!2\left(2\!-\!3\kappa_{o}^{2}\right)\kappa_{o}^{2}\ln\left(1\!+\frac{1}{\kappa_{o}}\right)\right]
	\]
	giving
	\[
	\gamma_{H}(0,\varphi_{0})\sqrt{\epsilon}\approx\frac{4\kappa_{o}}{\left(1+\kappa_{o}\right)^{2}}\exp\left\{ -\cos\left(4\varphi_{0}\right)\left[\frac{1}{2}-(2\kappa_{o}-1)\left(3\kappa_{o}^{2}-1\right)-2\left(2-3\kappa_{o}^{2}\right)\kappa_{o}^{2}\ln\left(1+\frac{1}{\kappa_{o}}\right)\right]\right\} .
	\]
This equation together with above dependence $\kappa_{o}(\varphi_{0})$
determine the low-temperature anisotropy of $H_{c2}$, which depends
on the azimuth angle $\varphi_{0}$. In particular, $\gamma_{H}(0,0)=1.13/\sqrt{\epsilon}$,
$\gamma_{H}(0,\pi/4)=0.9/\sqrt{\epsilon}$, and $\gamma_{H}(0,\pi/8)=1/\sqrt{\epsilon}$.

\subsubsection{Equatorial line node }\label{App:EqNodeLine}

Here we summarize the calculations for the case of the
order parameter with an equatorial nodal line for which the weight function
has the form $\Omega=\Omega_{0}\cos\theta$.
For the anisotropy at $T_{c}$ the integration in Eq.~\eqref{gam_lamTc} gives
\begin{equation}
\gamma^{2}(T_{c}) =\frac{\left(2+\epsilon\right)\sqrt{1-\epsilon}-3\sqrt{\epsilon}\arcsin\sqrt{1-\epsilon}}{2\epsilon\left[\left(1-4\epsilon\right)\sqrt{1-\epsilon}+3\epsilon^{3/2}\arcsin\sqrt{1-\epsilon}\right]}.\label{eq:gamTcEqNode}
\end{equation}
for $\epsilon<1$. This dependence is plotted in Fig.\ \ref{EqLineNode} by a red line.

The calculation of $\gamma_H(0)$ requires the normalization constant $\Omega_{0}$. The normalization condition
$\Omega_{0}^{2}\left\langle \cos^{2}\theta\right\rangle =1$ using
\[
\left\langle \cos^{2}\theta\right\rangle =\sqrt{\epsilon}\int_{0}^{1}\frac{x^{2}dx}{\left[1-\left(1-\epsilon\right)x^{2}\right]^{3/2}}=\frac{1}{1-\epsilon}-\frac{\sqrt{\epsilon}}{(1-\epsilon)^{3/2}}\arcsin\sqrt{1-\epsilon}
\]
gives Eq.~\eqref{Om}.

The integration in Eq. \eqref{eq:gamoEqNodeInt} can be performed analytically giving
\[
\left\langle \Omega^{2}\frac{v_{x}^{2}\!-\!\gamma_{o}^{2}v_{z}^{2}}{v_{x}^{2}\!+\!\gamma_{o}^{2}v_{z}^{2}}\right\rangle =\!1\!-\frac{2\gamma_{o}\epsilon^{3/2}}{\left(\sqrt{1\!-\!\epsilon}\!-\!\sqrt{\epsilon}\arcsin\sqrt{1\!-\!\epsilon}\right)\sqrt{1\!-\!\gamma_{o}^{2}\epsilon^{2}}}\left[\frac{\sqrt{(1\!-\!\epsilon)(1\!-\!\gamma_{o}^{2}\epsilon^{2})}}{\epsilon\left(1+\gamma_{o}\sqrt{\epsilon}\right)}+\ln\frac{\sqrt{\epsilon}\left(\sqrt{(1\!-\!\epsilon)\gamma_{o}^{2}\epsilon}+\sqrt{1\!-\!\gamma_{o}^{2}\epsilon^{2}}\right)}{\sqrt{1-\epsilon}\!+\!\sqrt{1\!-\!\gamma_{o}^{2}\epsilon^{2}}}\right].
\]
The optimal anisotropy factor corresponding to the vanishing of the right hand side can be evaluated numerically.

\subsubsection{Two polar point nodes }\label{App:PolarNodes}

Here we summarize the calculations for the case of the order parameter
with two polar point nodes for which the weight function has the form
$\Omega=\Omega_{0}\sin\theta$. For the anisotropy at $T_{c}$ the
integration in Eq.~\eqref{gam_lamTc} gives
\begin{equation}
\gamma^{2}(T_{c})=\frac{-\sqrt{\epsilon}\left(5-2\epsilon\right)\sqrt{1-\epsilon}+3\arcsin\sqrt{1-\epsilon}}{2\epsilon^{3/2}\left[\left(2+\epsilon\right)\sqrt{1-\epsilon}-3\sqrt{\epsilon}\arcsin\sqrt{1-\epsilon}\right]}.\label{eq:gamTcPolar}
\end{equation}
for $\epsilon<1$. This dependence is plotted in Fig.\ \ref{FigPolarNodes}
by a red line.

The integral in Eq. \eqref{eq:gamoPolar} can be taken analytically. In the case $\epsilon,\gamma_{o}^{2}\epsilon^{2}<1$ the result is
\[
\left\langle \Omega^{2}\frac{v_{x}^{2}\!-\!\gamma_{o}^{2}v_{z}^{2}}{v_{x}^{2}\!+\!\gamma_{o}^{2}v_{z}^{2}}\right\rangle \!=\!1\!-\!\frac{2\gamma_{o}\epsilon}{\left(-\sqrt{\epsilon(1\!-\!\epsilon)}\!+\!\arcsin\sqrt{1\!-\!\epsilon}\right)\sqrt{1\!-\!\gamma_{o}^{2}\epsilon^{2}}}\!
\left[-\frac{\sqrt{\left(1\!-\!\epsilon\right)\left(1\!-\!\gamma_{o}^{2}\epsilon^{2}\right)}}{1+\gamma_{o}\sqrt{\epsilon}}
\!-\!\ln\frac{\sqrt{\epsilon}\left(\gamma_{o}\sqrt{\epsilon}\sqrt{1\!-\!\epsilon}+\sqrt{1\!-\!\gamma_{o}^{2}\epsilon^{2}}\right)}{\sqrt{1\!-\!\epsilon}\!+\!\sqrt{1\!-\!\gamma_{o}^{2}\epsilon^{2}}}\right].
\]

\end{widetext}

\subsection{Two spheroidal Fermi surfaces}\label{App:TwoBands}

For the case of two spheroids, Eq. \eqref{eq:OptAnis} for the optimal
anisotropy factor becomes
\[
\sum_{\alpha=1,2}\zeta_{\alpha}\left\langle \frac{v_{x,\alpha}^{2}-\gamma_{o}^{2}v_{z,\alpha}^{2}}{v_{x,\alpha}^{2}+\gamma_{o}^{2}v_{z,\alpha}^{2}}\right\rangle _{\alpha}=0
\]
with the band weights $\zeta_{\alpha}=n_{\alpha}\Omega_{\alpha}^{2}$.
The averages in this equation can be evaluated as
\[
\left\langle \frac{v_{x,\alpha}^{2}-\gamma_{o}^{2}v_{z,\alpha}^{2}}{v_{x,\alpha}^{2}+\gamma_{o}^{2}v_{z,\alpha}^{2}}\right\rangle _{\alpha}=\frac{1-\sqrt{\epsilon_{\alpha}}\gamma_{o}}{1+\sqrt{\epsilon_{\alpha}}\gamma_{o}}.
\]
This gives the equation for $\gamma_{o}$
\[
\zeta_{1}\frac{1-\sqrt{\epsilon_{1}}\gamma_{o}}{1+\sqrt{\epsilon_{1}}\gamma_{o}}+\zeta_{2}\frac{1-\sqrt{\epsilon_{2}}\gamma_{o}}{1+\sqrt{\epsilon_{2}}\gamma_{o}}=0,
\]
which has the following solution
\begin{equation}
\gamma_{o} \! =\frac{\zeta_{-}}{2}\left(\frac{1}{\sqrt{\epsilon_{1}}}-\frac{1}{\sqrt{\epsilon_{2}}}\right)+\sqrt{\frac{\zeta_{-}^{2}}{4}\left(\frac{1}{\sqrt{\epsilon_{1}}}+\frac{1}{\sqrt{\epsilon_{2}}}\right)^{2}\!+\frac{4\zeta_{1}\zeta_{2}}{\sqrt{\epsilon_{1}\epsilon_{2}}}}\label{eq:TwoSpherVarAnis}
\end{equation}
with $\zeta_{-}=\zeta_{1}-\zeta_{2}$.

The low-temperature anisotropy factor is connected with $\gamma_{o}$
by Eq. \eqref{eq:AnisVar}. Computing the averages
\[
\left\langle \ln\left(\frac{v_{x,\alpha}^{2}+\gamma_{o}^{2}v_{z,\alpha}^{2}}{v_{ab,\alpha}^{2}}\right)\right\rangle _{\alpha}=2\left[\ln\left(1+\sqrt{\epsilon_{\alpha}}\gamma_{o}\right)-1\right],
\]
we finally obtain
\begin{equation}
\gamma_{H}(0)=\gamma_{o}\prod_{\alpha=1,2}\left(\frac{1+\sqrt{\epsilon_{\alpha}}\gamma_{o}}{2}\right)^{-2\zeta_{\alpha}}.\label{eq:TwoSphaAnisH0}
\end{equation}
This equation together with Eq. \eqref{eq:TwoSpherVarAnis} gives
the approximate variational estimate for the low-temperature anisotropy
of the upper critical field in the case of two spheroidal Fermi surfaces
with isotropic s-wave order parameters.

  \end{document}